\pdfoutput=1
\documentclass[12pt]{article}
\usepackage{amsmath}
\usepackage{amssymb}
\usepackage{graphicx}
\usepackage[mathscr]{euscript}
\usepackage{enumerate}
\usepackage{natbib}
\usepackage{caption}
\usepackage{epsfig,float}
\usepackage{booktabs,multirow}
\usepackage{hhline}
\usepackage{url} 
\usepackage{algorithm}
\usepackage{verbatim}
\usepackage{xcolor}
\usepackage[ansinew]{inputenc}
\usepackage[noend]{algpseudocode}

\newcommand{\blind}{0}

\def\rset{\mathbb R}

\def\S{\mathcal{S}}

\newcommand{\eqdef}{\ensuremath{\stackrel{\mathrm{def}}{=}}}

\def\M{\mathcal{M}}

\DeclareMathOperator*{\argmax}{\textsf{Argmax}}

\newcommand{\qed}{\nobreak \ifvmode \relax \else
      \ifdim\lastskip<1.5em \hskip-\lastskip
      \hskip1.5em plus0em minus0.8em \fi \nobreak
      \vrule height0.75em width0.5em depth0.25em\fi}

\makeatletter
\def\BState{\State\hskip-\ALG@thistlm}
\makeatother

\algblock{ParFor}{EndParFor}
\algnewcommand\algorithmicparfor{\textbf{parfor}}
\algnewcommand\algorithmicpardo{\textbf{do}}
\algnewcommand\algorithmicendparfor{\textbf{end\ parfor}}
\algrenewtext{ParFor}[1]{\algorithmicparfor\ #1\ \algorithmicpardo}
\algrenewtext{EndParFor}{\algorithmicendparfor}

\algtext*{EndParFor}%

\algnewcommand\algorithmicswitch{\textbf{switch}}
\algnewcommand\algorithmiccase{\textbf{case}}
\algnewcommand\algorithmicassert{\texttt{assert}}
\algnewcommand\Assert[1]{\State \algorithmicassert(#1)}%
\algdef{SE}[SWITCH]{Switch}{EndSwitch}[1]{\algorithmicswitch\ #1\ \algorithmicdo}{\algorithmicend\ \algorithmicswitch}%
\algdef{SE}[CASE]{Case}{EndCase}[1]{\algorithmiccase\ #1}{\algorithmicend\ \algorithmiccase}%
\algtext*{EndSwitch}%
\algtext*{EndCase}%



\begin{document}

\def\spacingset#1{\renewcommand{\baselinestretch}%
{#1}\small\normalsize} \spacingset{1}


\if0\blind
{
  \title{\bf Likelihood Inference for Large Scale Stochastic Blockmodels with Covariates based
	on a Divide-and-Conquer Parallelizable Algorithm with Communication}
  \author{Sandipan Roy, Yves Atchad\'e\thanks{The work of YA was partially supported by NSF award DMS-1228164}\hspace{.2cm}\\
    Department of Statistical Science, University College London\\
		Department of Statistics, University of Michigan\\
    and \\
   George Michailidis\thanks{\textit{The work of GM was partially supported by NSF awards DMS-1228164, DMS-1545277, IIS-1632730 and by NIH award 5R01GM114029-02}}\\
    Department of Statistics \& Informatics Institute, University of Florida}
  \maketitle
} \fi

\if1\blind
{
  \bigskip
  \bigskip
  \bigskip
  \begin{center}
    {\LARGE\bf Likelihood Inference for Large Stochastic Blockmodels with Covariates}
\end{center}
  \medskip
} \fi

\bigskip
\begin{abstract}
We consider a stochastic blockmodel equipped with node covariate information, that is helpful in analyzing social network data.
The key objective is to obtain maximum likelihood estimates of the model parameters. For this task, we devise a fast, scalable Monte Carlo EM type algorithm based on case-control approximation of the log-likelihood coupled with a subsampling approach. A key feature of the proposed algorithm is its parallelizability, by processing portions of the data on several cores, while leveraging communication of key statistics across the cores during each iteration of the algorithm. The performance of the algorithm is evaluated on synthetic data sets and compared with competing methods for blockmodel parameter estimation.  We also illustrate the model on data from a Facebook derived social network enhanced with node covariate information.
\end{abstract}

\noindent%
\vfill

\newpage
\spacingset{1.45} 
\section{Introduction}
\label{sec:intro}

There has been a lot of recent work in modeling network data, primarily driven by novel applications in social network analysis,
molecular biology, public health, etc. A common feature of network data in numerous applications is the presence of {\em community
structure}, which means that a subset of nodes exhibits higher degree of connectivity amongst themselves than the remaining
nodes in the network. The problem of community detection has been extensively studied in the statistics and networks literature,
and various approaches proposed, including spectral clustering (\cite{white2005spectral}, \cite{rohe2011spectral} etc.), likelihood based methods (\cite{airoldi2008mixed}, \cite{amini2013}, \cite{nowicki2001estimation} etc.), and modularity based techniques (\cite{girvan2002community}), as well as approaches inspired by statistical physics principles (\cite{fortunato2010community}).

For likelihood based methods, a popular generative statistical model used is the \textit{Stochastic Block Model} (SBM)
(\cite{holland1983stochastic}). Edges in this model are generated at random with probabilities corresponding to
entries of an inter-community probability matrix, which in turn leads to community structures in the network. However, on many applications, the network data are complemented either by node-specific or edge-specific covariates. Some of the available work in the literature focuses on node covariates
for the SBM (or some variant of it) (\citet{tallberg2004bayesian, mariadassou2010uncovering, choi2012stochastic, airoldi2008mixed}), while other papers focus on edge-specific covariates (\cite{hoff2002latent, mariadassou2010uncovering,choi2012stochastic}).


The objective of this work is to obtain maximum likelihood estimates (MLE) of the model parameters in large scale SBMs with covariates. This is a challenging computational problem, since the latent structure of the model requires an EM-type algorithm to obtain the estimates. It is known (\citet{snijders1997estimation, handcock2007model}) that for a network of size $n$, each EM update requires $O(n^2)$ computations, an expensive calculation for large networks. Further, one also needs $O(nK)$ calculations to obtain the community memberships, which could also prove to be a computationally expensive step for large $n$, especially if the number of communities $K$ scales with $n$.

\citet{amini2013} provided a pseudo-likelihood method for community detection in large sparse networks, which can be used for fast parameter estimation in a regular SBM, but it is not readily applicable to settings where the SBM has also covariates. The recent work \cite{ma2017exploration} deals with large scale likelihood-based inference for networks, but focuses on latent space models. Hence, there is a need to scale up likelihood-based inference for large SBMs with covariates. The goal of this work is to fill that gap. To deal with the computational problem we develop a divide-and-conquer parallelizable algorithm that can take advantage of multi-processor computers. The algorithm allows communication between the processors during its iterations. As shown in Section \ref{sec:methods}, this communication step improves estimation accuracy,
while creating little extra computational overhead, compared to a straightforward divide-and-conquer parallelizable algorithm. We believe that
such an algorithm is particularly beneficial for inference purposes when the data exhibit intricate dependencies, such as in an SBM. To boost performance, the proposed algorithm is enhanced with a case-control approximation of the log-likelihood.

The remainder of the paper is organized as follows:
In Section \ref{sec:sbm}, we describe the general K-class SBM with covariates and present a Monte-Carlo EM for SBM with covariates in Section \ref{mcem:covsec}. In Section \ref{sec:ccapprox}, we give a general overview of the case-control approximation used for faster computation of the log-likelihood in large network data and also discuss the specific approximation employed for the log-likelihood in SBMs.
In Section \ref{par:algodesc}, we describe two generic parallel schemes in estimating the parameters of the model, in Section \ref{simresults}, we provide numerical evidence on simulated data regarding the performance of the proposed algorithm together with comparisons with two existing latent space models utilizing covariate information viz. (1) an additive and mixed effects model focusing on dyadic networks (AMEN) (\citet{hoff2005bilinear,hoff2015dyadic}) and (2) a latent position cluster model using Variational Bayes implementation (VBLPCM) (\citet{salter2013variational}). We conclude with a real data application involving Facebook networks of US colleges with a specific number of covariates in Section \ref{sec:application}.


\section{Modeling Framework and a Scalable Algorithm}
\label{sec:methods}

\subsection{A SBM with covariates}
\label{sec:sbm}

Suppose that we have a $0-1$ symmetric adjacency matrix $A=((a_{ij}))\in\rset^{n\times n}$, where $a_{ii}=0$. It corresponds to an undirected graph with nodes $\{1,\ldots,n\}$, where there is an edge between nodes $(i,j)$, if $a_{ij}=1$. Suppose that in addition to the adjacency matrix $A$, we observe some symmetric covariates $X(i,j)=X(j,i)\in\rset^p$ on each pair of nodes $(i,j)$ on the graph that influence the formation of the graph. In such cases, it is naturally appealing to extend the basic SBM to include the covariate information. Let $Z=(Z_1,\ldots,Z_n)$ denote the group membership of the $n$ nodes. We assume that $Z_i\in\{1,\ldots,K\}$, and that the $Z_i$'s are independent random variables with a multinomial distribution with probabilities $\pi=(\pi_1,\ldots,\pi_K)$. We assume that given $Z$, the random variables $\{a_{ij},\;1\leq i<j\leq n\}$ are conditionally independent Bernoulli random variables, and
\begin{equation}
\label{sbm:covariate}
a_{ij}\sim\mbox{Ber}(P_{ij}),\;\mbox{ where } \log\frac{P_{ij}}{1-P_{ij}}=\theta_{Z_iZ_j}+\beta^TX(i,j),\;\;\;1\leq i<j\leq n,
\end{equation}
with $\theta\in\mathbb{R}^{K\times K}$ being a symmetric matrix.  The parameter of the model is $\xi \equiv (\theta,\beta,\pi)\in\Xi\eqdef\rset^{K\times K}\times \rset^p\times \Delta$, where $\Delta$ is the set of probability distributions on $\{1,\ldots,K\}$. For some convenience in the notation we shall henceforth write $\xi$ to denote the parameter set $(\theta,\beta,\pi)$. A recent paper by \cite{latouche2015goodness} also considered a logistic model for random graphs with covariate information. Their goal was to assess the goodness of fit of the model, where the network structure is captured by a graphon component. To overcome the intractability of the graphon function,
the original model is approximated by a sequence of models involving a blockstructure. An instance of that approximation corresponds to the proposed model, but the direct objectives of the two works are rather different.

The log-likelihood of the posited model for the observed data is given by
\begin{equation}
\label{ll:obdata}
\log\int_{\mathcal{Z}} L(\theta,\beta,\pi|A,z) dz,
\end{equation}
where $\mathcal{Z} = \{1,\ldots,K\}^n$, and $L(\xi|A,z) = L(\theta,\beta,\pi|A,z)$ is the complete data likelihood given by
\begin{equation}
\label{lcomp:cov}
L(\xi|A,z) =\prod_{i<j}\left(\frac{\mathrm{e}^{\theta_{z_iz_j}+\beta^TX(i,j)}}{1+\mathrm{e}^{\theta_{z_iz_j}+\beta^TX(i,j)}}\right)^{a_{ij}}\left(\frac{1}{1+\mathrm{e}^{\theta_{z_iz_j}+\beta^TX(i,j)}}\right)^{1-a_{ij}}\prod_{i=1}^n \pi_{z_i}.
\end{equation} 
Although $\mathcal{Z}$ is a discrete set, we write it as an integral with respect to a counting measure for notational convenience. When $n$ is large, obtaining the maximum-likelihood estimate (MLE) $$(\hat{\theta},\hat{\beta},\hat{\pi})=\displaystyle\argmax_{(\theta,\beta,\pi)\in\Xi}\;\log\int_{\mathcal{Z}} L(\theta,\beta,\pi|A,z) dz$$ is a difficult computational problem. 
We describe below a Monte Carlo EM (MCEM) implementation for parameter estimation of the proposed SBM with covariates.

 \subsection{Monte Carlo EM for SBM with Covariates}
\label{mcem:covsec}
As mentioned in the introductory section, since direct computation of the log-likelihood or its gradient is intractable, estimating SBMs is a nontrivial computational task, especially for large size networks. The MCEM algorithm (\cite{wei1990monte}) is a natural algorithm to tackle this problem. Let $p(\cdot\vert \xi, A)$ denotes the posterior distribution on $\mathcal{Z}$ of the latent variables $z=(z_1,\ldots,z_n)$ given parameter $\xi = (\theta,\beta,\pi)$ and data $A$. More precisely,
\begin{equation}\label{eq:cond:dist:latent:z}
p(z\vert \xi, A) \propto \prod_{i<j}\left(\frac{\mathrm{e}^{\theta_{z_iz_j}+\beta^TX(i,j)}}{1+\mathrm{e}^{\theta_{z_iz_j}+\beta^TX(i,j)}}\right)^{a_{ij}}\left(\frac{1}{1+\mathrm{e}^{\theta_{z_iz_j}+\beta^TX(i,j)}}\right)^{1-a_{ij}} \prod_{i=1}^n \pi_{z_i}.\end{equation}
We assume that we have available a Markov kernel $\mathcal{K}_{\xi, A}$ on $\mathcal{Z}$ with invariant distribution $p(\cdot\vert\xi, A)$ that we can use to generate MCMC draws from $p(\cdot\vert \xi, A)$. In all our simulations below a Gibbs sampler (\cite{robert2013monte}) is used for that purpose.   We now present the main steps of the MCEM algorithm for a SBM with covariates. 

\begin{algorithm}[!ht]
\caption{Basic Monte Carlo EM}\label{basic:mcem}
\begin{itemize}
\item Initialize $\xi_0 = (\theta_0,\beta_0,\pi_0)$
\item At the $r$-th iteration, given working estimate $\xi_r = (\theta_r,\beta_r,\pi_r)$, do the following two steps.
\begin{description}
\item [(E-step)] Generate a Markov sequence\footnotemark{} 
$(z^{(1)}_{r+1}, \ldots, z_{r+1}^{(M_r)})$, using the Markov kernel $\mathcal{K}_{\xi_r,A}$ with invariant distribution $p(\cdot|\xi_r,A)$. Use this Monte Carlo sample to derive the approximate Q-function
\begin{equation}
\label{Q-function:EM covariate}
\widehat{Q}\left(\xi;\xi_r\right)=\frac{1}{M_r}\displaystyle\sum_{m=1}^{M_r}\log L\left(\theta,\beta,\pi|A,z^{(m)}_{r+1}\right).
\end{equation}
\item [(M-step)] Maximize the approximate Q-function to obtain a new estimates:
\[\xi_{r+1} = \left(\theta_{r+1},\beta_{r+1},\pi_{r+1}\right)=\displaystyle\argmax_{\xi\in\Xi} \ \ \widehat{Q}\left(\xi;\xi_r\right).\]
\end{description}
\item Repeat the above two steps for $r=1,2,\ldots$ until convergence.
\end{itemize}
\end{algorithm}
\footnotetext{We draw the initial state $z_{r+1}^{(0)}$ using spectral clustering with perturbation (\citet{amini2013}). However other choices are possible.}

Because the Monte Carlo samples are allowed to change with the iterations, the MCEM algorithm described above generates a non-homogeneous Markov chain with sequence of transition kernels $\{\M_r,\;r\geq 1\}$, where $\M_r(\xi_r,A;\cdot)$ denote the conditional distribution of $\xi_{r+1}$ given $(\xi_0,\ldots,\xi_r)$. We made explicit the dependence of these transition kernels on the dataset $A$. This notation will come handy later on as we run the same algorithm on different datasets. Using this notation, the MCEM algorithm can be succinctly presented as follows: choose some initial estimate $\xi_0\in\Xi$; for $r=1,\ldots$, draw
\[\xi_{r+1}\vert (\xi_0,\ldots,\xi_r) \sim \M_r(\xi_r,A,\cdot).\]
This representation is very convenient, and helps providing a clear description of the main algorithm introduced below.

The $r$-th iteration of the MCEM algorithm outlined above requires $\mathcal{O}(n^2M_r)$ calculations\footnote{A more precise cost estimate is $\mathcal{O}(dn^2M_r)$, where $d$ is the number of covariates. However here we assume that $d$ is much small compared to $n$ and $M_r$.}, where $M_r$ is the number of Monte Carlo samples used at iteration $r$ and $n$ denotes the number of network nodes. Note that since MCMC is used for the Monte Carlo approximation, large values of $M_r$ are typically needed to obtain reasonably good estimates\footnote{In fact, since the mixing of the MCMC algorithm would typically depend on the size of $\mathcal{Z}=\{1,\ldots,K\}^n$ (and hence on $n$), how large $M_r$ should be to obtain a reasonably good Monte Carlo approximation in the E-step depends in an increasing fashion on $n$.}. This demonstrates that obtaining  the MLE for the posited model becomes computationally expensive as the size of the network $n$ grows.  The main bottleneck is the computation of the complete data log-likelihood
\begin{equation}
\label{loglik:compldata}
\log L(\xi|A,z)=\sum_{i<j}\left[a_{ij}\left(\theta_{z_iz_j}+\beta^TX(i,j)\right) - \log\left(1+\mathrm{e}^{\theta_{z_iz_j}+\beta^TX(i,j)}\right)\right] + \sum_{i=1}^n \log\pi_{z_i}.
\end{equation} 
We use the case-control approximation (\citet{raftery2012fast}) to obtain a fast approximation of the log-likelihood $\log L\left(\xi|A,z\right)$. A general overview of this approximation and the specific implementation for the model under consideration
are provided in the next section.

\subsection{Case-Control Approximation in Monte Carlo EM}
\label{sec:ccapprox}
The main idea of case-control approximations comes from cohort studies,  where the presence of case subjects is relatively rare compared to that of control subjects (for more details see \citet{breslow1996statistics, breslow1982statistical}). In a network context, if its topology is relative sparse (there are a number of tightly connected communities, but there do not exist too many connections between members of different communities), then the number of edges (cases) is relatively small compared  to the absence of edges (controls). Then, the sum in Equation~\eqref{loglik:compldata} consists mostly of terms with $a_{ij}=0$ and therefore fast computation of the likelihood through case-control approximation (\citet{raftery2012fast}) becomes attractive.
Specifically, splitting the individual by group, we can express the log-likelihood as
\begin{equation}
\label{loglik:ccapprox}
\ell(\theta,\beta,\pi\vert A,z) \equiv \log L(\theta,\beta,\pi|A,z)= \frac{1}{2}\displaystyle\sum_{k=1}^K \sum_{i:\;z_i=k} \left[\ell_i\left(\theta,\beta|A,z\right) + \log\pi_k\right]
\end{equation}
where
\begin{align*}
\begin{split}
\ell_i\left(\theta,\beta|A,z\right) &\equiv\displaystyle\sum_{j\neq i}\left\{a_{ij}\left(\theta_{z_iz_j}+\beta^T X(i,j)\right)-\log\left(1+\mathrm{e}^{\theta_{z_iz_j}+\beta^T X(i,j)}\right)\right\}\\
    &=\displaystyle\sum_{j\neq i,a_{ij}=1}\left\{\left(\theta_{z_iz_j}+\beta^T X(i,j)\right)-\log\left(1+\mathrm{e}^{\theta_{z_iz_j}+\beta^T X(i,j)}\right)\right\}\\ &-\displaystyle\sum_{j\neq i,a_{ij}=0}\log\left(1+\mathrm{e}^{\theta_{z_iz_j}+\beta^T X(i,j)}\right)\\
		&=\ell_{i,1} + \ell_{i,0},
\end{split}
\end{align*}
where
\[\ell_{i,0} \equiv -\displaystyle\sum_{j\neq i,a_{ij}=0}\log\left(1+\mathrm{e}^{\theta_{z_iz_j}+\beta^T X(i,j)}\right).\]
Given a node $i$, with $z_i=k$, we set $\mathcal{N}_{i,0}=\{j\neq i:\; a_{ij}=0\}$, and  $\mathcal{N}_{i,g,0}=\{j\neq i:\; z_j=g,\; a_{ij}=0\}$ for some group index $g$. Using these notations we further split the term $\ell_{i,0}$ as
\[ \ell_{i,0} = -\sum_{g=1}^K \displaystyle\sum_{j\in \mathcal{N}_{i,g,0}}\log\left(1+\mathrm{e}^{\theta_{kg}+\beta^T X(i,j)}\right).\]
Let $\S_{i,g}$ denotes a randomly selected\footnote{We do an equal-probability random selection with replacement. If $m_0\geq |\mathcal{N}_{i,g,0}|$ an exhaustive sampling is done} subset of size $m_0$ from the set $\mathcal{N}_{i,g,0}$. Following the case control approximation, we approximate  the term $\ell_{i,0}$ by
\[\tilde{\ell}_{i,0}=-\sum_{g=1}^K \frac{N_{i,g,0}}{m_0}\displaystyle\sum_{J\in\S_{i,g,0}}\log\left(1+\mathrm{e}^{\theta_{kg}+\beta^T X(i,J)}\right),\]
where $N_{i,g,0} = |\mathcal{N}_{i,g,0}|$ is the cardinality of $\mathcal{N}_{i,g,0}$. Note that $\tilde{\ell}_{i,0}$ is an unbiased Monte Carlo estimate of $\ell_{i,0}$. Hence 
\[\tilde \ell_i(\theta,\beta|A,z) = \ell_{i,1} + \tilde \ell_{i,0}\]
is an unbiased Monte Carlo estimate of $\ell_i\left(\theta,\beta\vert A,z\right)$, and 
\begin{equation}
\label{approx:compll}
\tilde \ell(\theta,\beta,\pi|A,z)=\frac{1}{2}\displaystyle\sum_{k=1}^K\displaystyle\sum_{i:z_i=k} \left[\tilde \ell_i(\theta,\beta|A,z) + \log\pi_k\right],
\end{equation}
is an unbiased estimator of the log-likelihood. Hence, one can use a relatively small sample $m_0K$ to obtain an unbiased and fast approximation of the complete log-likelihood. The variance decays like $O(1/(Km_0))$. In this work we have used a simple random sampling scheme. Other sampling schemes developed with variance reduction in mind can be used as well, and this include stratified case-control sampling (\citet{raftery2012fast}), local case-control subsampling (\citet{fithian2014local}). However these schemes come with additional computational costs.

The case-control approximation leads to an approximation of the conditional distribution of the latent variables $z$ given by
\[\tilde p(z\vert A,\xi) \propto e^{\tilde \ell(\theta,\beta,\pi|A,z)},\]
which replaces (\ref{eq:cond:dist:latent:z}). As with the basic MCEM algorithm, we assume that we can design, for any $\xi\in\Xi$, a Markov kernel $\widetilde{\mathcal{K}}_{\xi}$ on $\mathcal{Z}$ with invariant distribution $\tilde p(\cdot\vert A,\xi)$ that can be easily implemented. In our implementation a Gibbs sampler is used. We thus obtain a new (case-control approximation based) Monte Carlo EM algorithm.

\begin{algorithm}[H]
\caption{Case-Control Monte Carlo EM}\label{cc:mcem}
\begin{itemize}
\item Initialize $\xi_0 = (\theta_0,\beta_0,\pi_0)$
\item At the $r$-th iteration, given working estimate $\xi_r = (\theta_r,\beta_r,\pi_r)$, do the following two steps.
\begin{enumerate}
\item \label{approx Estep:MCEM}
Generate a Markov chain $(z^{(1)}_{r+1}, \ldots, z_{r+1}^{(M_r)})$ with transition kernel $\widetilde{\mathcal{K}}_{\xi_r,A}$ and invariant distribution $\tilde p(\cdot |\xi_r,A)$.  Use this Monte Carlo sample to form
\begin{equation}
\label{Q-function:ccEM covariate}
\widetilde{Q}\left(\xi;\xi_r\right)=\frac{1}{M_r}\displaystyle\sum_{m=1}^{M_r} \tilde{\ell}\left(\theta,\beta,\pi|A,z^{(m)}_{r+1}\right).
\end{equation}
\item \label{approx Mstep:MCEM}
Compute the  new estimate
\[\xi_{r+1} = \displaystyle\argmax_{\xi\in\Xi} \ \ \widetilde{Q}\left(\xi;\xi_r\right).\]
\end{enumerate}
\item Repeat the above two steps for $r=1,2,\ldots$ until convergence.
\end{itemize}
\end{algorithm}

As with the MCEM algorithm, we will compactly represent the Case-Control MCEM algorithm as a non-homogeneous Markov chain with a sequence of transition kernels $\{\widetilde \M_r,\;r\geq 1\}$.

In conclusion, using the case-control approximation reduces the computational cost of every EM iteration from $O(n^2M_r)$ to $O(Km_0nM_r)$, where $Km_0\ll n$ is the case-control sample size. In our simulations, we choose $m_0 = \lambda r$, where $\lambda$ is the average node degree of the network, and $r$ is the global case-to-control rate.

\section{Parallel implementation by sub-sampling}
\label{par:algodesc}
The Case-Control Monte Carlo EM described in \textbf{Algorithm} \ref{cc:mcem} could still be expensive to use for very large networks. We propose a parallel implementation of the algorithm to further reduce the computational cost. The main idea is to draw several sub-adjacency matrices that are processed in parallel on different machines. The computational cost is hence further reduced since the case-control MCEM algorithm is now applied on smaller adjacency matrices. The novelty of our approach resides in the proposed parallelization scheme.


Parallelizable algorithms have recently become popular for very large-scale statistical optimization problems; for example \cite{nedic2009distributed,ram2010distributed,johansson2009randomized,duchi2012dual} considered distributed computation for minimizing a sum of convex objective functions. For solving the corresponding optimization problem, they considered subgradient methods in a distributed setting. \cite{zhang2013communication} considered a straightforward divide and conquer strategy and show a reduction in the mean squared error for the parameter vector minimizing the population risk under the parallel implementation compared to a serial method. Their applications include large scale linear regression, gradient based optimization, etc. The simple divide and conquer strategy of parallel implementation has also been studied for some classification and estimation problems by \cite{mcdonald2009efficient, mcdonald2010distributed}, as well as for certain stochastic approximation methods by \cite{zinkevich2010parallelized} and by \cite{recht2011hogwild} for a variant of parallelizable stochastic gradient descent. \citet{dekel2012optimal} considered a gradient based online prediction algorithm in a distributed setting, while \citet{agarwal2011distributed} considered optimization in an asynchronous distributed setting based on delayed stochastic gradient information.

Most of the literature outlined above has focused on the divide and conquer (with no communication) strategy. However this strategy works only in cases where the random subsamples from the dataset produce unbiased estimates of the gradient of the objective function. Because of the inherent heterogeneity of network data, this property does not hold for the SBM. Indeed, fitting the SBM on a randomly selected sub-adjacency matrix can lead to sharply biased estimate of the parameter\footnote{Consider for instance the extreme case where all the nodes selected belong to the same community.}. We introduce a parallelization scheme where running estimates are shared between the machines to help mitigate the bias. 

Suppose that we have $T$ machines to be used to fit the SBM. Let $\{A^{(u)},\;u=1,\ldots,T\}$ be a set of $T$ randomly and independently selected sub-adjacency matrices from $A$, where $A^{(u)}\in\{0,1\}^{n_0\times n_0}$. These sub-matrices can be drawn in many different ways. Here we proceed as follows. Given an initial clustering of the nodes (by spectral clustering with perturbation (\citet{amini2013})) into $K$ groups, we draw the sub-matrix $A^{(u)}$ by randomly selecting $\lfloor n_0/K\rfloor$ nodes with replacement from each of the $K$ groups. The sub-matrix $A^{(u)}$ is then assigned (and sent to) machine $u$. A divide and conquer approach to fitting the SBM consists in running, without any further communication between machines, the case-control MCEM algorithm for $R$ iterations on each machine: for each $u=1,\ldots,T$
\[\xi^{(u)}_r \vert(\xi^{(u)}_0,\ldots,\xi_{r-1}^{(u)})  \sim \widetilde\M_{r-1}(\xi_{r-1}^{(u)},A^{(u)};\cdot),\;\;\;r=1,\ldots,R.\]
Then we estimate $\xi$ by
\[\frac{1}{T}\sum_{u=1}^T \xi_R^{(u)}.\]
This plain divide and conquer algorithm is summarized in \textbf{Algorithm} \ref{parallel:wcomm}.

To mitigate the potential bias due to using sub-adjacency matrices, we allow the machines to exchange their running estimates after each iteration. More precisely, after the $r$-th iteration a master processor collects all the running estimates $\{\xi_r^{(i)},\;1\leq i\leq T\}$ (where $T$ is the number of slave processors), and then send estimate $\xi_r^{(1)}$ to processor $2$, $\xi_r^{(2)}$ to processor $3$, etc... and send $\xi_r^{(T)}$ to processor $1$. In this fashion, after $T$ iterations or more, each running estimate has been updated based on all available sub-adjacency matrices, and this helps mitigate any potential bias induced by the selected sub-matrices. The algorithm is summarized in \textbf{Algorithm} \ref{parallel:comm}.  
The computational cost is similar to the no-communication scheme, but we now have the additional cost of communication which on most shared-memory computing architecture would be relatively small. At the end of the $R$-th iteration, we estimate $\xi$ by
\[\frac{1}{T}\sum_{u=1}^T \xi_R^{(u)}.\]

\begin{algorithm}[H]
\caption{Parallel Case-Control Monte Carlo EM without Communication}\label{parallel:wcomm}
\begin{algorithmic}[1]
\Require Adjacency matrix $A\in\rset^{n\times n}$, random subsamples $\left\{A^{(i)}\right\}_{i=1}^T\in\mathbb{R}^{n_0\times n_0}$, Number of machines $T$, Number of iterations $R$.
\Ensure $\bar{\xi}_R=\frac{1}{T}\displaystyle\sum_{i=1}^T\xi_R^{(i)}$
\BState For each machine $i$ initialize $\xi_0^{(i)}=\left(\theta_0^{(i)},\beta_0^{(i)},\pi_0^{(i)}\right)$
\ParFor {$i=1$ to $T$} (for each machine)
\For {$r=1$ to $R$}, draw
\State $\xi_r^{(i)}\vert (\xi_0^{(i)},\ldots,\xi_{r-1}^{(i)}) \sim \widetilde\M_{r-1}\left(\xi_{r-1}^{(i)},A^{(i)};\cdot\right)$.
\EndFor
\State \textbf{end}
\EndParFor
\State \textbf{end}
\end{algorithmic}
\end{algorithm}

\begin{algorithm}
\caption{Parallel Case-Control Monte Carlo EM with Communication}\label{parallel:comm}
\begin{algorithmic}[1]
\Require Adjacency matrix $A\in\rset^{n\times n}$, random subsamples $\left\{A^{(i)}\right\}_{i=1}^T\in\mathbb{R}^{n_0\times n_0}$, Number of machines $T$, Number of iterations $R$.
\Ensure $\bar{\xi}_R=\frac{1}{T}\displaystyle\sum_{i=1}^T\xi_R^{(i)}$
\BState For each machine $i$ initialize $\xi_0^{(i)}=\left(\theta_0^{(i)},\beta_0^{(i)},\pi_0^{(i)}\right)$
\For {$r=1$ to $R$} (for each iteration)
\ParFor {$i=1$ to $T$} (parallel computation),
\State $\check\xi_r^{(i)}\vert (\xi_0^{(i)},\ldots,\xi_{r-1}^{(i)})\sim\widetilde\M_{r-1}\left(\xi_{r-1}^{(i)},A^{(i)};\cdot\right)$.
\EndParFor
\State \textbf{end}
\State Set $\xi = \check\xi_r^{(T)}$.
\For { $i=2$ to $T$} (exchange of running estimates)
\State $\xi_r^{(i)} = \check\xi_r^{(i-1)}$.
\EndFor
\State \textbf{end}
\State $\xi_r^{(1)} = \xi$.
\EndFor
\State \textbf{end}
\end{algorithmic}
\end{algorithm}

\section{Performance evaluation}
\label{simresults}
We compare the proposed algorithm (Algorithm \ref{parallel:comm}) with Algorithm \ref{parallel:wcomm} (non-communication case-control MCEM), and with the baseline MCEM algorithm using the full data (Algorithm \ref{basic:mcem}). We also include in the comparison the pseudo-likelihood method of \cite{amini2013}. We simulate observations from the SBM given in Equation~\eqref{sbm:covariate} as follows.  We fix the number of communities to $K$=3, and the network size to $n=1000$. We generate the latent membership vector $z=\left(z_1,z_2,\ldots,z_n\right)$ as independent random variables from a Multinomial distribution with parameter $\pi$. We experiment with two different class probabilities for the 3 communities, viz. $\pi=(1/3, 1/3, 1/3)^{\prime}$ (balanced community size) and $\pi=(0.5, 0.3, 0.2)^{\prime}$ (unbalanced community size).

We vary two intrinsic quantities related to the network, namely the out-in-ratio (OIR) (denoted $\mu$) and the average degree (denoted $\lambda$). The OIR $\mu$ (\citet{decelle2011asymptotic}) is the ratio of the number of links between members in different communities to the number of links between members of same communities. We vary $\mu$ as $(0.04, 0.08, 0.2)$ which we term as \textit{low} OIR, \textit{medium} OIR and \textit{high} OIR, respectively. The average degree $\lambda$ is defined as $n$ times the ratio of the total number of links present in the network to the total number of possible pairwise connections (that is ${n\choose 2}$). We vary $\lambda$ in the set $(4, 8, 14)$, which we term as \textit{low}, \textit{medium} and \textit{high} degree regimes, respectively. Using $\mu$ and $\lambda$, and following \cite{amini2013}, we generate the link probability matrix $\theta\in \mathbb{R}^{3\times 3}$ as follows
\[\theta = \frac{\lambda}{(n-1)\pi^T\theta^{(0)}\pi}\theta^{(0)},\;\;\mbox{ where }\;\; \theta^{(0)} = \left(\begin{tabular}{ccc} $\mu$ & 1 & 1 \\ 1 & $\mu$ & 1 \\ 1 & 1 & $\mu$ \end{tabular}\right).\]

We set the number of covariates to $p=3$ and the regression coefficients $\beta$ to $(1,-2,1)$. For each pair of nodes $(i,j)$, its covariates are generated by drawing $p$ independent $\text{Ber}(0,1)$ random variables. And we obtain the probability of a link between any two individuals $i$ and $j$ in the network as
\[P_{ij}=\frac{\exp(\theta_{z_iz_j}+\beta^TX(i,j))}{1+\exp(\theta_{z_iz_j}+\beta^TX(i,j))}.\]
Given the latent membership vector $z$, we then draw the entries of an adjacency matrix $A=((a_{ij}))_{n\times n}$ as
\[a_{ij}\stackrel{\text{ind}}{\sim}\mbox{Ber}(P_{ij})\mbox{ $i,j=1,2,\ldots,n$ }\]

We evaluate the algorithms using the mean squared error (MSE) of the parameters $\pi,\theta,\beta$, and a measure of recovery of the latent node labels obtained by computing the Normalized Mutual Information (NMI) between the recovered clustering and the true clustering (\citet{amini2013}). The Normalized Mutual Information between two sets of clusters $C$ and $C^{\prime}$ is defined
\[\text{NMI}=\frac{I(C,C^{\prime})}{H(C)+H(C^{\prime})}\] 
where $H(\cdot)$ is the entropy function and $I(\cdot,\cdot)$ is the mutual information between the two sets of clusters. We have $\text{NMI}\in [0,1]$, and the two sets of clusters are similar if NMI is close to 1.

For all algorithms we initialize $\xi$ as follows. We initialize the node labels $z$ using spectral clustering with perturbations (\cite{amini2013}), that we subsequently use to initialize $\pi_0$ as 
\[\pi_{0k} = \frac{1}{n}\sum_{i=1}^n \textbf{1}(z_i=k).\]
We initialize $\theta$ by
\[\theta_0(a,b) = \frac{\sum_{i\neq j} A_{ij}\textbf{1}(z_{0i} = a)\textbf{1}(z_{0j}=b)}{\sum_{i\neq j} \textbf{1}(z_{0i} = a)\textbf{1}(z_{0j}=b)},\]
and we initialize the regression parameter $\beta$ by fitting a logistic regression using the binary entries of the adjacency $A(i,j)$ are responses and $X(i,j)$ are covariates.

For the case-control algorithms we employ a global case-to-control rate $r=7$, so that the case-control sample sizes are set to $\lambda r = 7\lambda$. We also choose the subsample size to be $\lfloor\frac{n_0}{K}\rfloor=50$ from each group where $K$ is the number of groups.  All the simulations were replicated  30 times.

We first illustrate the statistical and computational performance of the parallelizable MCEM algorithm {\em with} and {\em without} communication on a small network of size $n=100$ with $K=3$ communities and latent class probability vector $\pi=\left(1/3, 1/3, 1/3\right)^{\prime}$). The results are depicted in Table \ref{toyexample:methods}.

\begin{center}
\captionof{table}{Estimation Errors and NMI Values (standard errors are in parenthesis) for Balanced Community Size with Varying out-in-ratio (OIR)}
\label{toyexample:methods}
\begin{tabular}[!ht]{|p{5.0cm}|p{1.8cm}|p{1.8cm}|p{1.8cm}|p{1.8cm}|p{1.4cm}|}
\hline
Methods & estimation error($\theta$) & estimation error($\pi$) & estimation error ($\beta$) & NMI(z) & Time\\
\hhline{------}
\text{MCEM on Full Data} & 0.1721 & 0.1812 & 0.1578 & 0.6184 (\textcolor{red}{0.0134}) & 57.80\text{sec}\\	\hline
\text{Parallel Communication} & 0.1921 & 0.2061 & 0.1643 & 0.5901 (\textcolor{red}{0.0157}) & 12.58\text{sec}\\	\hline
\text{Parallel Non-communication} & 0.2202 & 0.2141 & 0.1793 & 0.6107 (\textcolor{red}{0.0387}) & 12.46\text{sec}\\ \hline
\end{tabular}
\end{center}

It can be seen that both versions of the parallel MCEM algorithm are almost five times faster than the serial one; further, the 
communications based variant is ~10\% inferior in terms of statistical accuracy on all parameters of interest, while the performance of the non-communication one is ~20\% worse than the full MCEM algorithm. Similar performance of the communications variant, albeit with larger estimation gains has been observed in many other settings of the problem.\\Tables \ref{perftab1result} and \ref{perftab2result} 
depict the results when $\text{OIR}$ is varied from \textit{low} to \textit{high}. In Tables \ref{perftab1result} and \ref{perftab2result}, the average degree is kept at 8. Along with the MSE we also report the bias of parameters $\left(\pi,\theta,\beta\right)$ in the parenthesis in Tables \ref{perftab1result}-\ref{perftab4result}. One can observe that the MSE for different parameters for parallel MCEM with communication is only around ~10\% worse than the corresponding values for MCEM on the full data. On the other hand, the non-communicative parallel version could be more than ~50\% worse than the MCEM on the full data and could possibly be even worse in the high OIR regime for unbalanced communities. In Table 2 for $\text{OIR}=0.2$ one can observe that the bias reduction in the parameter estimates is between 60-90\% whereas gain in the NMI is only about 3\% (colored red in Table \ref{perftab1result}).\\Tables \ref{perftab3result} and \ref{perftab4result} show the performance of the three different methods when the average degre $\lambda$ is varied from the \textit{low} to the \textit{high} regime. The OIR is kept at 0.04 in both Tables. As before, we observe significant improvements in MSE for the communications version over its non-communications counterpart, with the gain being even higher for smaller $\lambda$ values compared to the higher ones. In Table \ref{perftab4result} for $\lambda=4$ one can observe that the bias reduction in the parameter estimates is between 62-90\% whereas NMI increases from non-communication setting to communication one only by 2\% (colored red in Table \ref{perftab3result}). The similar trend in bias reduction compared to the NMI value, albeit with different percentages of reduction are observable in other settings of OIR and $\lambda$. Further, the performance of parallel MCEM with communication is close to the level of performance of MCEM on the full data over different values of $\lambda$. The NMI values for the communications version is around ~4\% better than the non-communications one.\\

\begin{figure}[ht]
\centering
\begin{tabular}{ccc}
\epsfig{file=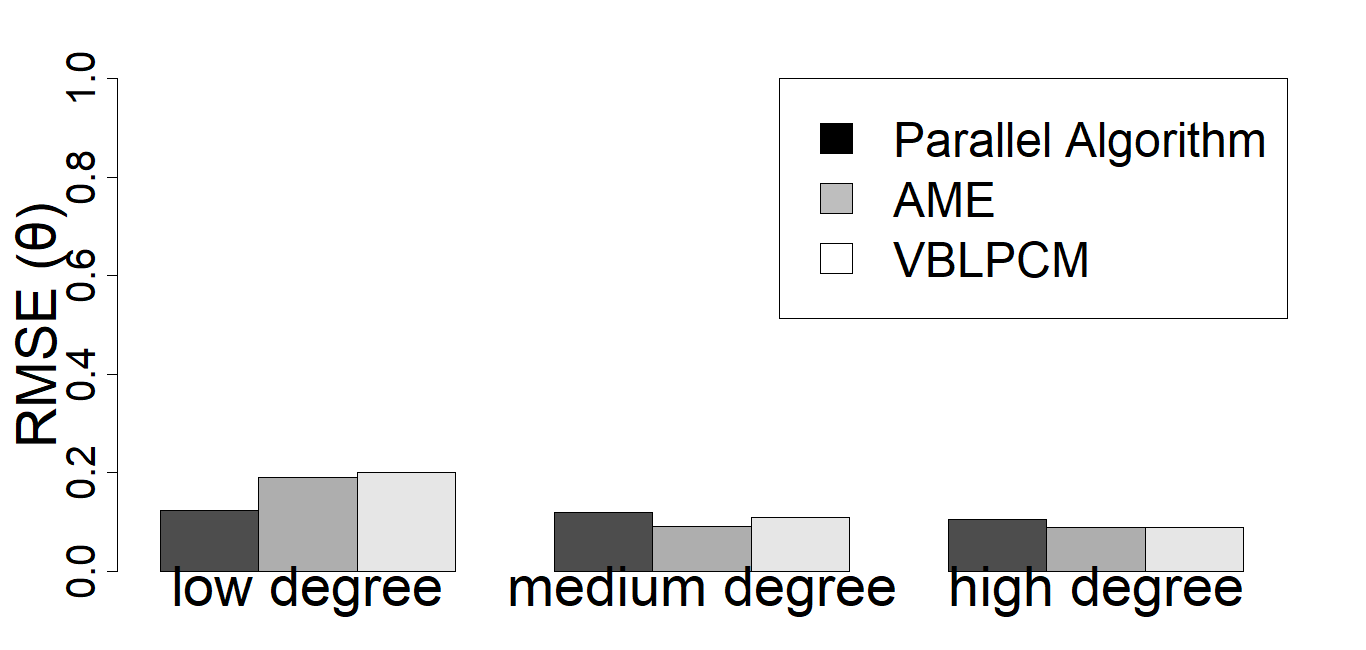,width=0.3\linewidth,clip=} & 
\epsfig{file=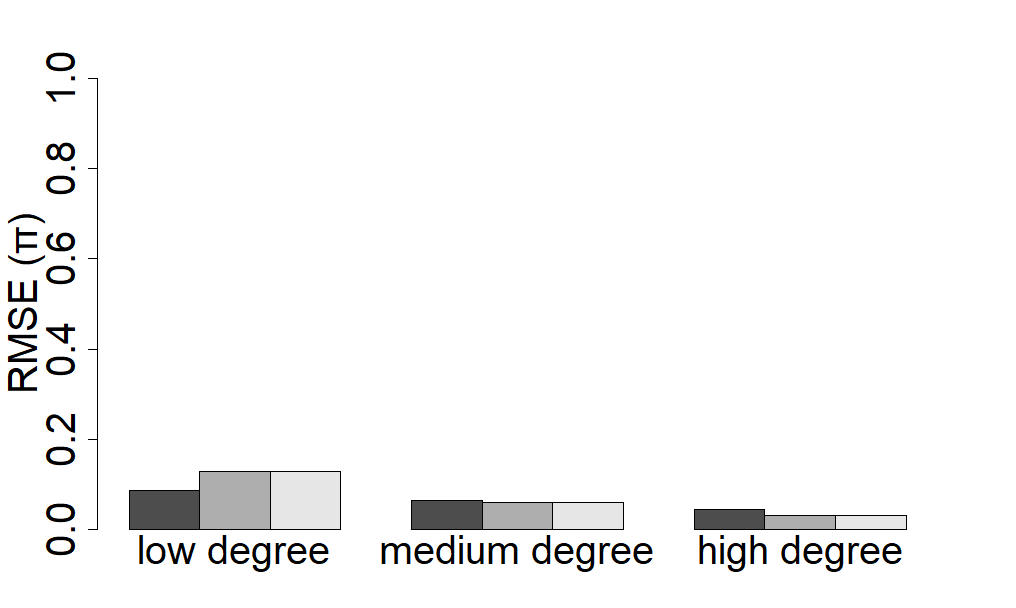,width=0.3\linewidth,clip=} &
\epsfig{file=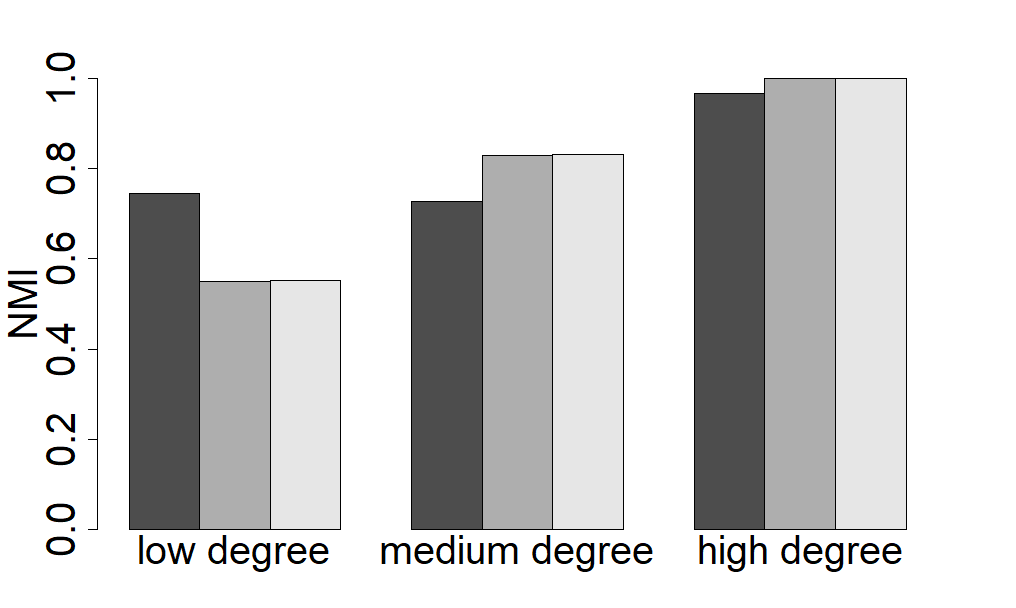,width=0.3\linewidth,clip=} \\
\epsfig{file=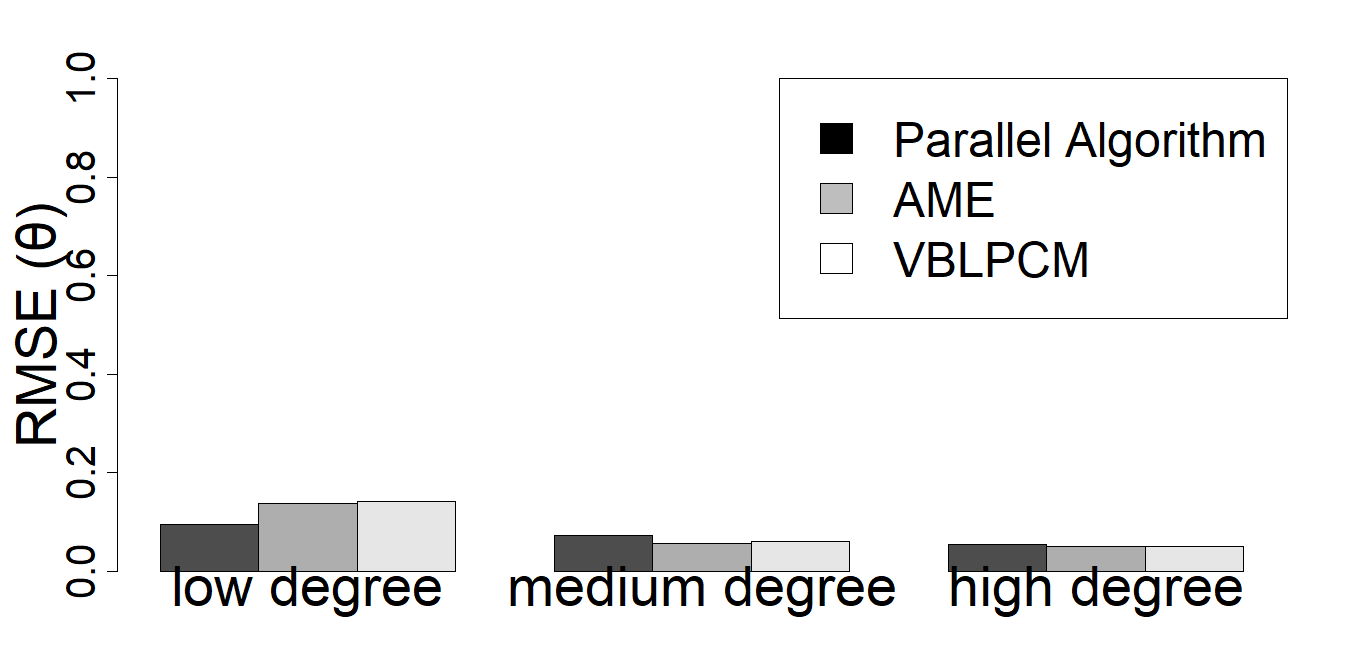,width=0.3\linewidth,clip=} &
\epsfig{file=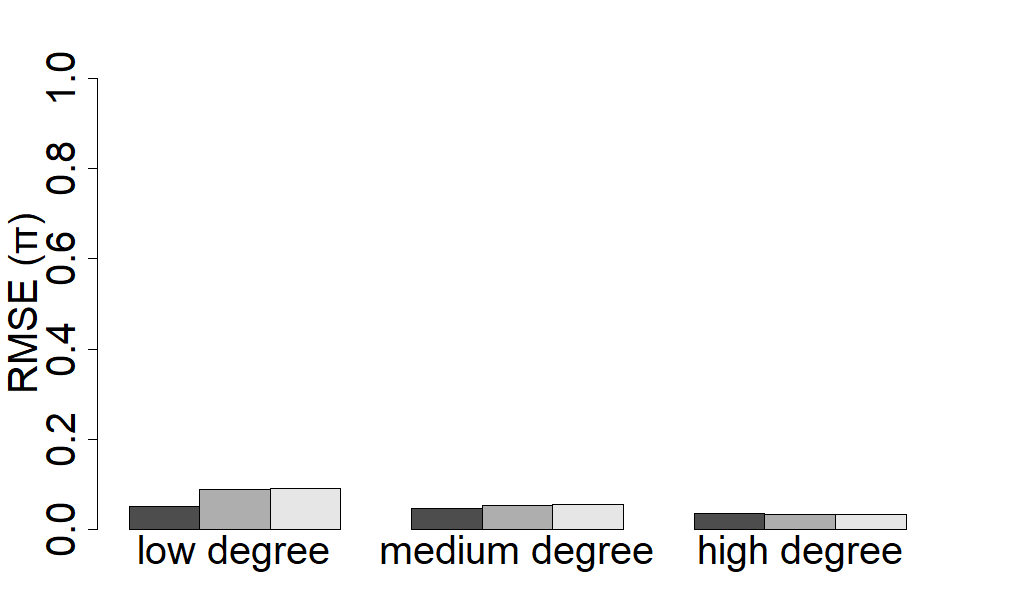,width=0.3\linewidth,clip=} &
\epsfig{file=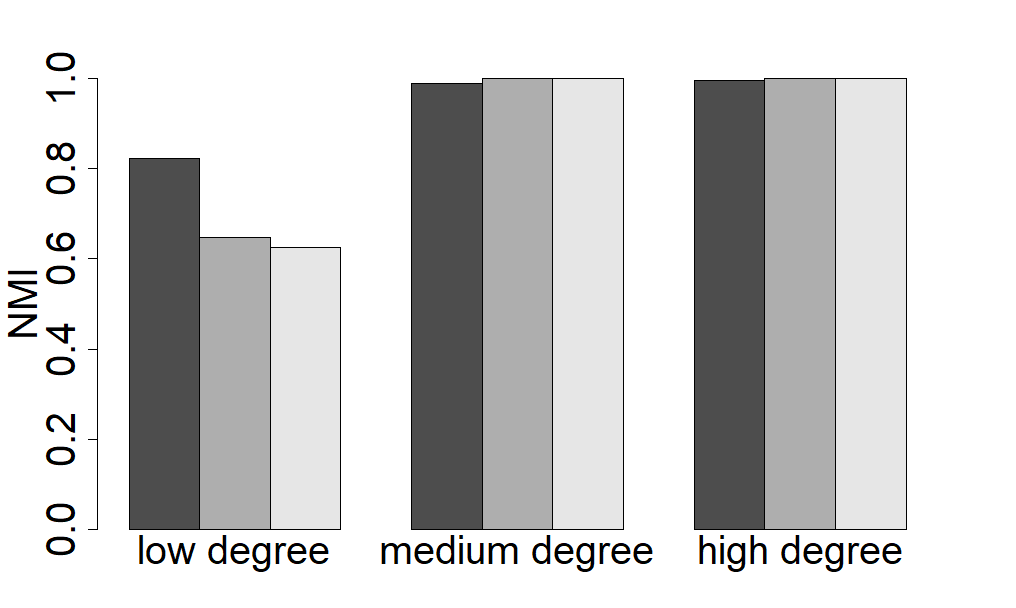,width=0.3\linewidth,clip=} \\
\end{tabular}
\caption{Comparison of \textbf{Algorithm} \ref{parallel:comm} to the additive and mixed effect linear models in networks (AMEN) and a variational Bayes implementation of latent position cluster model (VBLPCM) for parameter estimation in Stochastic Blockmodels for low, medium and high degree networks, respectively. Top row corresponds to the unbalanced, while bottom row to the balanced community size case, respectively.}
\label{fig3}
\end{figure}

\begin{figure}[ht]
\centering
\begin{tabular}{ccc}
\epsfig{file=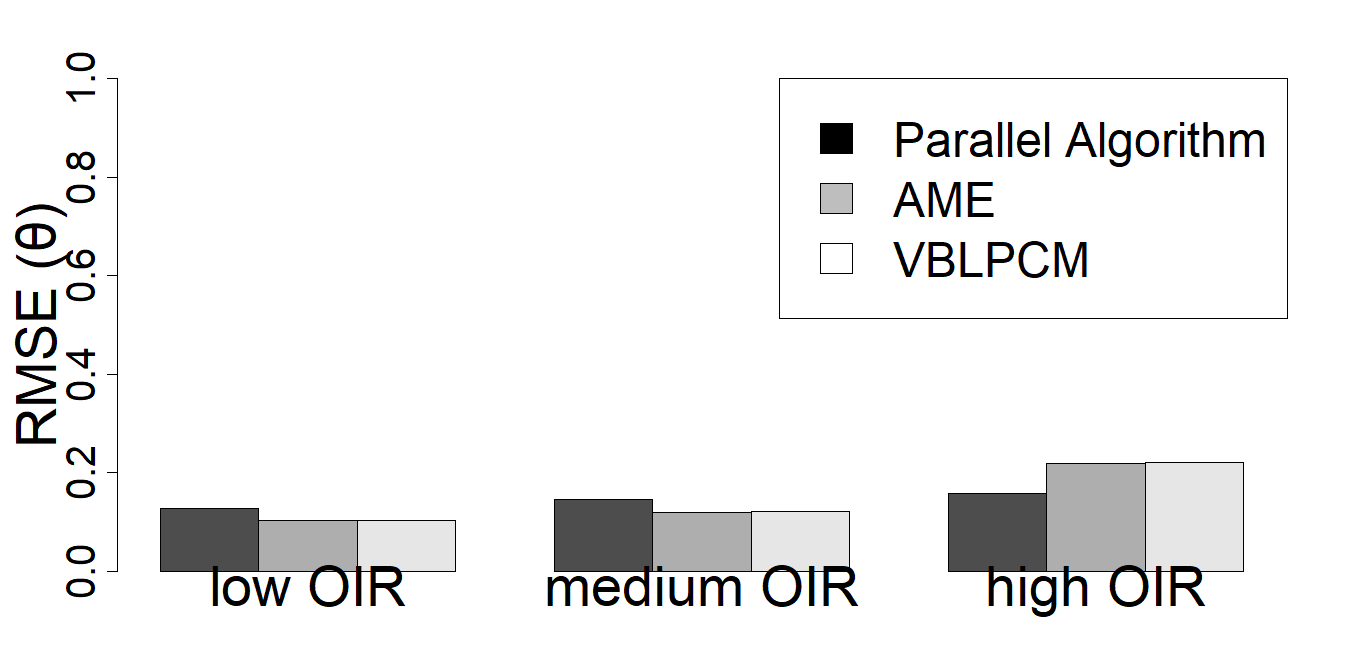,width=0.3\linewidth,clip=} & 
\epsfig{file=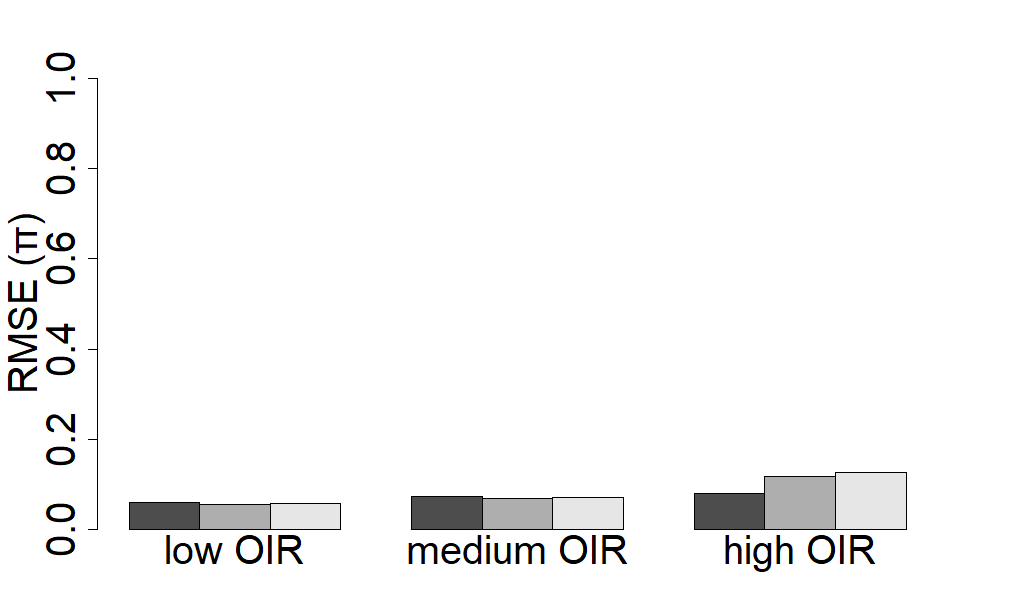,width=0.3\linewidth,clip=} &
\epsfig{file=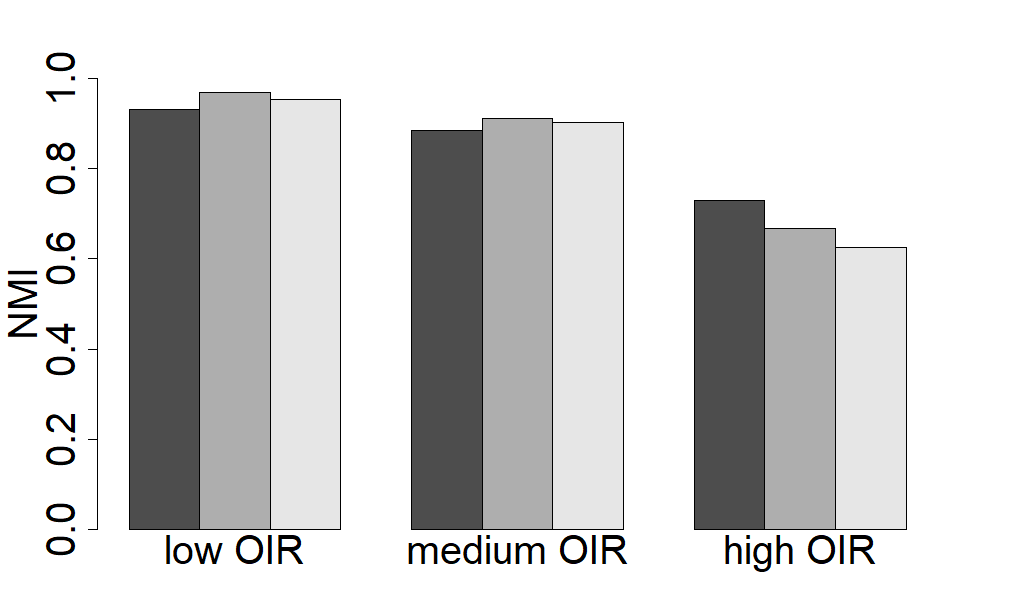,width=0.3\linewidth,clip=} \\
\epsfig{file=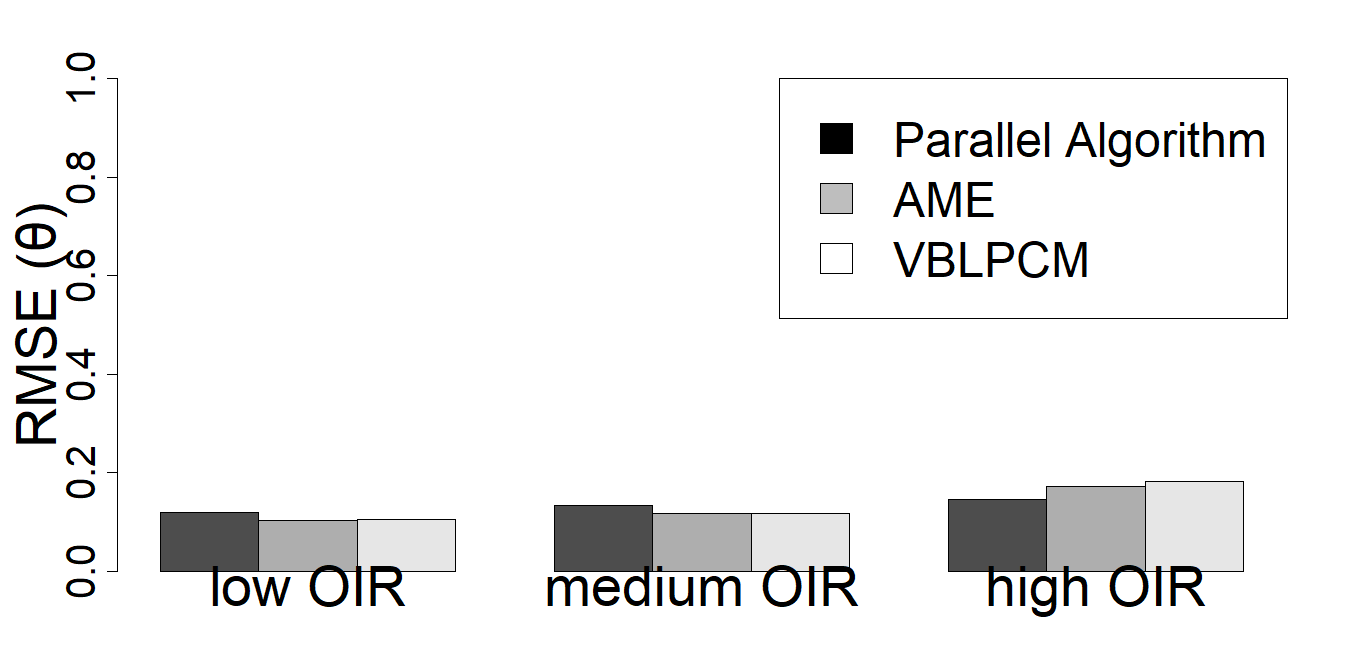,width=0.3\linewidth,clip=} &
\epsfig{file=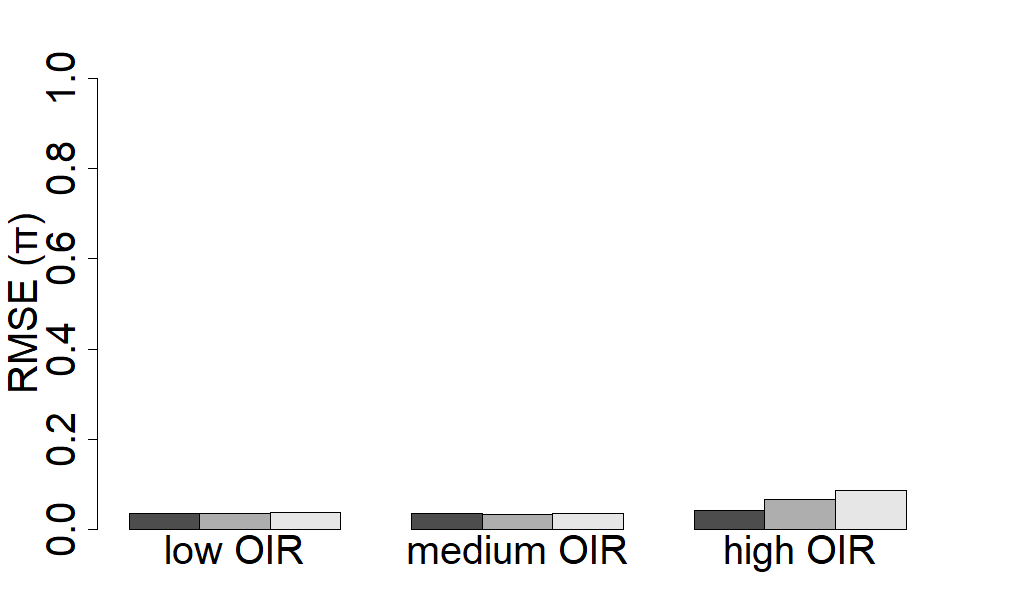,width=0.3\linewidth,clip=} &
\epsfig{file=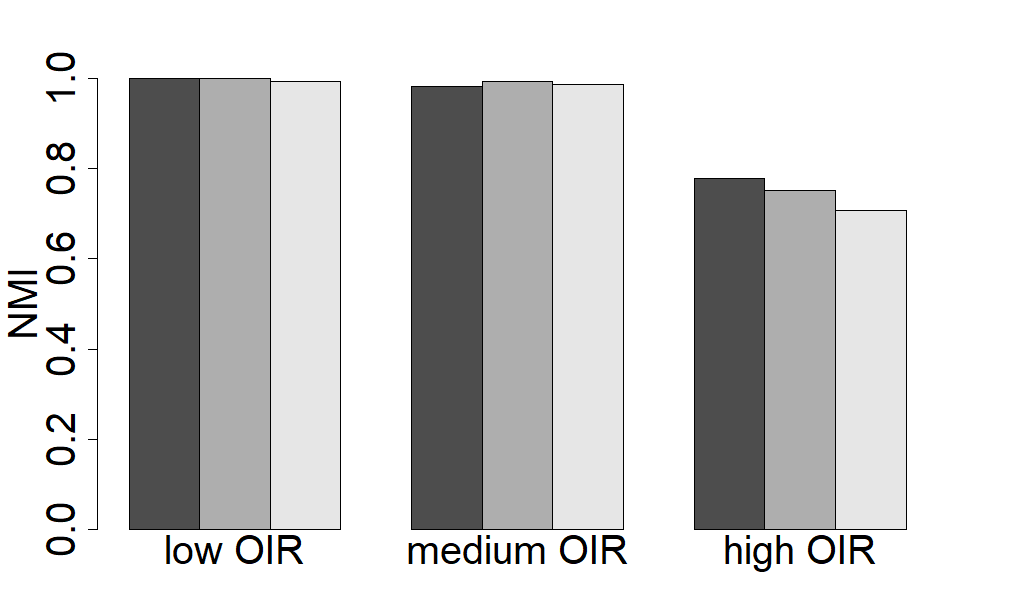,width=0.3\linewidth,clip=} \\
\end{tabular}
\caption{Comparison of \textbf{Algorithm} \ref{parallel:comm} to the additive and mixed effect linear models in networks (AMEN) and a variational Bayes implementation of latent position cluster model (VBLPCM) for parameter estimation in Stochastic Blockmodels for low, medium and high OIR networks, respectively. Top row corresponds to the unbalanced, while bottom row to the balanced community size case, respectively.}
\label{fig4}
\end{figure}

We also compare the proposed modeling approach and \textbf{Algorithm} \ref{parallel:comm} to two other models in the literature- (1) an additive and mixed effects model focusing on dyadic networks (AMEN) (\citet{hoff2005bilinear,hoff2015dyadic}) and (2) a latent position cluster model using Variational Bayes implementation (VBLPCM) (\citet{salter2013variational}). As before, we use two different settings- balanced and unbalanced community size and make the comparison in the bar diagrams given in Figures~\ref{fig3} and ~\ref{fig4}, respectively. Also in one case, we keep the average degree $\lambda$ fixed at 8 and vary OIR as $(0.04, 0.08, 0.2)$, while on another occasion we fix OIR at 0.04 and vary $\lambda$ as $(4, 8, 14)$. To compare the performance of our parallel communication algorithm to AMEN and VBLPCM with respect to community detection, we use bar diagrams of the NMI values under the settings described above. Based on the results depicted in Figures~\ref{fig3} and \ref{fig4}, we observe that both AMEN and VBLPCM tend to 
exhibit a slightly better performance in terms of NMI values and RMSE of parameter estimates when OIR is low (assortative network structure) or medium and $\lambda$ is medium or high. Our parallel algorithm tends to perform significantly better than both AMEN and VBLPCM when OIR is high and $\lambda$ is low. In fact, gains for AMEN and VBLPCM in terms of performance over \textbf{Algorithm} \ref{parallel:comm} in the mentioned settings are less compared to the gain of \textbf{Algorithm} \ref{parallel:comm} over its competitors in high OIR (disassortative network structure) and low $\lambda$ (sparse) settings. The simulation studies do convey the fact that for sparse networks and in cases where communities have high interactions (many real world networks have one or both of these features) amongst their member nodes, \textbf{Algorithm} \ref{parallel:comm} exhibits a superior performance compared to AMEN or VBLPCM for likelihood based inference in SBMs.

\begin{small}
\begin{center}
\captionof{table}{Comparison of performance of three different methods for $\lambda=8$, $n=1000$, $K=3$ and balanced community size with varying OIR (bias of the estimates are given in parentheses)\label{perftab1result}}  
\begin{tabular}{l c c c c c}  
\hline\hline                       
 OIR & Methods & est.err($\pi$) & est.err($\theta$) & est.err($\beta$) & NMI
\\ [0.5ex]   
\hline
\multirow{3}{*}{0.04} & \text{MCEM on Full Data} & 0.0313 & 0.0893 & 0.0185 & 1.0000\\
                    & \text{Parallel Communication} & 0.0340 (0.0020) & 0.0987 (0.0049) & 0.0232 (0.0016) & 1.0000\\
										& \text{Parallel Non-communication} & 0.0483 (0.0039) & 0.1194 (0.0078) & 0.0433 (0.0035) & 0.9000\\ \hline
\multirow{3}{*}{0.08} & \text{MCEM on Full Data} & 0.0321 & 0.0916 & 0.0228 & 0.9876\\
                    & \text{Parallel Communication} & 0.0349 (0.0024) & 0.1042 (0.0060) & 0.0320 (0.0020) & 0.9830\\
										& \text{Parallel Non-communication} & 0.0568 (0.0043) & 0.1377 (0.0104) & 0.0549 (0.0039) & 0.8939\\ \hline
\multirow{3}{*}{\textcolor{red}{0.2}} & \text{MCEM on Full Data} & 0.0385 & 0.0988 & 0.0378 & 0.7916\\
                    & \textcolor{red}{Parallel Communication} & \textcolor{red}{0.0406 (0.0029)} & \textcolor{red}{0.1061 (0.0079)} & \textcolor{red}{0.0476 (0.0036)} & \textcolor{red}{0.7796}\\
										& \textcolor{red}{Parallel Non-communication} & \textcolor{red}{0.0617 (0.0358)} & \textcolor{red}{0.1459 (0.0671)} & \textcolor{red}{0.0701 (0.0091)} & \textcolor{red}{0.7534}\\ \hline
										
\end{tabular}
\end{center}

\begin{center}
\captionof{table}{Comparison of performance of three different methods for $\lambda=8$, $n=1000$, $K=3$ and unbalanced community size with varying OIR (bias of the estimates are given in parentheses)\label{perftab2result}}  
\begin{tabular}{l c c c c c}  
\hline\hline                       
 OIR & Methods & est.err($\pi$) & est.err($\theta$) & est.err($\beta$) & NMI
\\ [0.5ex]   
\hline
\multirow{3}{*}{0.04} & \text{MCEM on Full Data} & 0.0511 & 0.0879 & 0.0412 & 0.9510\\
                    & \text{Parallel Communication} & 0.0604 (0.0036) & 0.0937 (0.0047) & 0.0644 (0.0045) & 0.9327\\
										& \text{Parallel Non-communication} & 0.0782 (0.0051) & 0.1185 (0.0077) & 0.0750 (0.0053) & 0.8681\\ \hline
\multirow{3}{*}{0.08} & \text{MCEM on Full Data} & 0.0589 & 0.0933 & 0.0612 & 0.9054\\
                    & \text{Parallel Communication} & 0.0736 (0.0048) & 0.1048 (0.0068) & 0.0732 (0.0051) & 0.8852\\
										& \text{Parallel Non-communication} & 0.0874 (0.0065) & 0.1253 (0.0125) & 0.0867 (0.0069) & 0.8428\\ \hline
\multirow{3}{*}{0.2}  & \text{MCEM on Full Data} & 0.0657 & 0.1041 & 0.0804 & 0.8251\\
                    & \text{Parallel Communication} & 0.0803 (0.0058) & 0.1187 (0.0088) & 0.0954 (0.0072) & 0.7896\\
										& \text{Parallel Non-communication} & 0.1010 (0.0586) & 0.1503 (0.0691) & 0.1309 (0.0170) & 0.7314\\ \hline
										
\end{tabular}
\end{center}

\begin{center}
\captionof{table}{Comparison of performance of three different methods for $OIR=0.04$, $n=1000$, $K=3$ and balanced community size with varying $\lambda$ (bias of the estimates are given in parentheses)\label{perftab3result}}  
\begin{tabular}{l c c c c c}  
\hline\hline                       
 $\lambda$ & Methods & est.err($\pi$) & est.err($\theta$) & est.err($\beta$) & NMI
\\ [0.5ex]   
\hline
\multirow{3}{*}{\textcolor{red}{4}} & \text{MCEM on Full Data} & 0.0467 & 0.0885 & 0.0455 & 0.8532\\
                    & \textcolor{red}{Parallel Communication} & \textcolor{red}{0.0508 (0.0037)} & \textcolor{red}{0.0948 (0.0070)} & \textcolor{red}{0.0516 (0.0049)} & \textcolor{red}{0.8240}\\
										& \textcolor{red}{Parallel Non-	communication} & \textcolor{red}{0.0664 (0.0385)} & \textcolor{red}{0.1343 (0.0698)} & \textcolor{red}{0.0724 (0.0145)} & \textcolor{red}{0.8084}\\ \hline
\multirow{3}{*}{8} & \text{MCEM on Full Data} & 0.0389 & 0.0703 & 0.0393 & 0.9976\\
                    & \text{Parallel Communication} & 0.0451 (0.0028) & 0.0721 (0.0053) & 0.0487 (0.0034) & 0.9889\\
										& \text{Parallel Non-communication} & 0.0604 (0.0054) & 0.0925 (0.0148) & 0.0613 (0.0061) & 0.9670\\ \hline
\multirow{3}{*}{14} & \text{MCEM on Full Data} & 0.0302 & 0.0508 & 0.0297 & 1.0000\\
                    & \text{Parallel Communication} & 0.0340 (0.0020) & 0.0540 (0.0035) & 0.0354 (0.0025) & 0.9968\\
										& \text{Parallel Non-communication} & 0.0515 (0.0031) & 0.0805 (0.0056) & 0.0575 (0.0046) & 0.9856\\ \hline
										
\end{tabular}
\end{center}

\begin{center}
\captionof{table}{Comparison of performance of three different methods for $OIR=0.04$, $n=1000$, $K=3$ and unbalanced community size with varying $\lambda$ (bias of the estimates are given in parentheses)\label{perftab4result}}  
\begin{tabular}{l c c c c c}  
\hline\hline                       
 $\lambda$ & Methods & est.err($\pi$) & est.err($\theta$) & est.err($\beta$) & NMI
\\ [0.5ex]   
\hline
\multirow{3}{*}{4} & \text{MCEM on Full Data} & 0.0778 & 0.1189 & 0.0651 & 0.7832\\
                    & \text{Parallel Communication} & 0.0853 (0.0061) & 0.1244 (0.0092) & 0.0706 (0.0053) & 0.7447\\
										& \text{Parallel Non-communication} & 0.1052 (0.0610) & 0.1605 (0.0738) & 0.1082 (0.0141) & 0.7192\\ \hline
\multirow{3}{*}{8} & \text{MCEM on Full Data} & 0.0554 & 0.1087 & 0.0543 & 0.8982\\
                    & \text{Parallel Communication} & 0.0628 (0.0041) & 0.1186 (0.0071) & 0.0612 (0.0043) & 0.8681\\
										& \text{Parallel Non-communication} & 0.0815 (0.0059) & 0.1419 (0.0114) & 0.0811 (0.0081) & 0.8337\\ \hline
\multirow{3}{*}{14} & \text{MCEM on Full Data} & 0.0368 & 0.0974 & 0.0410 & 0.9889\\
                    & \text{Parallel Communication} & 0.0433 (0.0026) & 0.1047 (0.0052) & 0.0478 (0.0033) & 0.9668\\
										& \text{Parallel Non-communication} & 0.0575 (0.0040) & 0.1286 (0.0077) & 0.0695 (0.0049) & 0.9334\\ \hline
										
\end{tabular}
\end{center}
\end{small}

\section{Application to Collegiate Facebook Data}
\label{sec:application}
\begin{figure}
\includegraphics[scale=0.7]{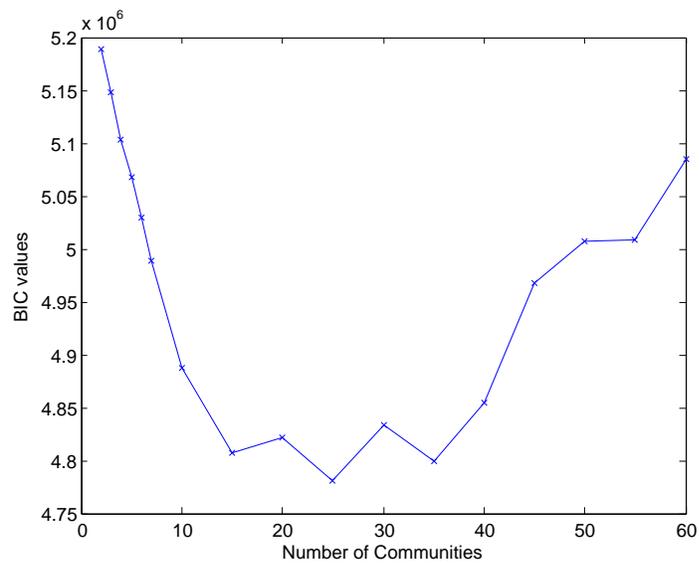}
\vspace{-15em}
\caption{Choice of the number of clusters (communities) in the Rice University Dataset. Plot of BIC values over possible number of clusters in the Dataset.}
\label{fig_bic}
\end{figure}

We use the proposed model to analyze a publicly available social network data set. The data come from {\tt https://archive.org/details/oxford-2005-facebook-matrix} that contains the social structure of Facebook friendship networks at one hundred American colleges and universities at a single point in time. This data set was analyzed by \cite{traud2012social} . The focus of their study was to illustrate how the relative importance of different characteristics of individuals vary across different institutions. They examine the influence of the common attributes at the dyad level in terms of assortativity coefficients and regression models. We on the other hand pick a data set corresponding to a particular university and show the performance of our algorithm and compare the clusters obtained from it with the ones obtained in case of fitting an SBM without covariates.\\We examine the Rice University data set from the list of one hundred American colleges and universities and use our K-class SBM with and without covariates to identify group/community structures in the data set. We examine the role of the user attributes- dorm/house number, gender and class year along with the latent structure.\\ Dorm/house number is a multi-category variable taking values as 202, 203, 204 etc., gender is a binary ($\left\{0,1\right\}$) variable and class year is a integer valued variable (e.g. ``2004'', ``2005'', ``2006'' etc.). We evaluate the performance of \textbf{Algorithm} \ref{parallel:comm} fitted to SBM with covariate viz. model~\eqref{sbm:covariate}.\\ There are some missing values in the data set although it is only around 5\%. Since the network size is 4087 which is large enough, we discard the missing value cases. We also consider the covariate values only between year 2004 to 2010. Further, we drop those nodes with degree less than or equal to 1. After this initial cleaning up, the adjacency matrix is of order $3160\times 3160$. We choose number of communities $K=20$. The choice of the number of communities is made by employing Bayesian Information Criterion (BIC) where the observed data likelihood is computed by path sampling (\citet{gelman1998simulating}). The corresponding plot is given in Figure~\ref{fig_bic} where the possible number of communities are plotted along the horizontal axis and the BIC values along the vertical one. 

\begin{figure}[ht]
\centering
\begin{tabular}{ccc}
\hspace{-40pt}\epsfig{file=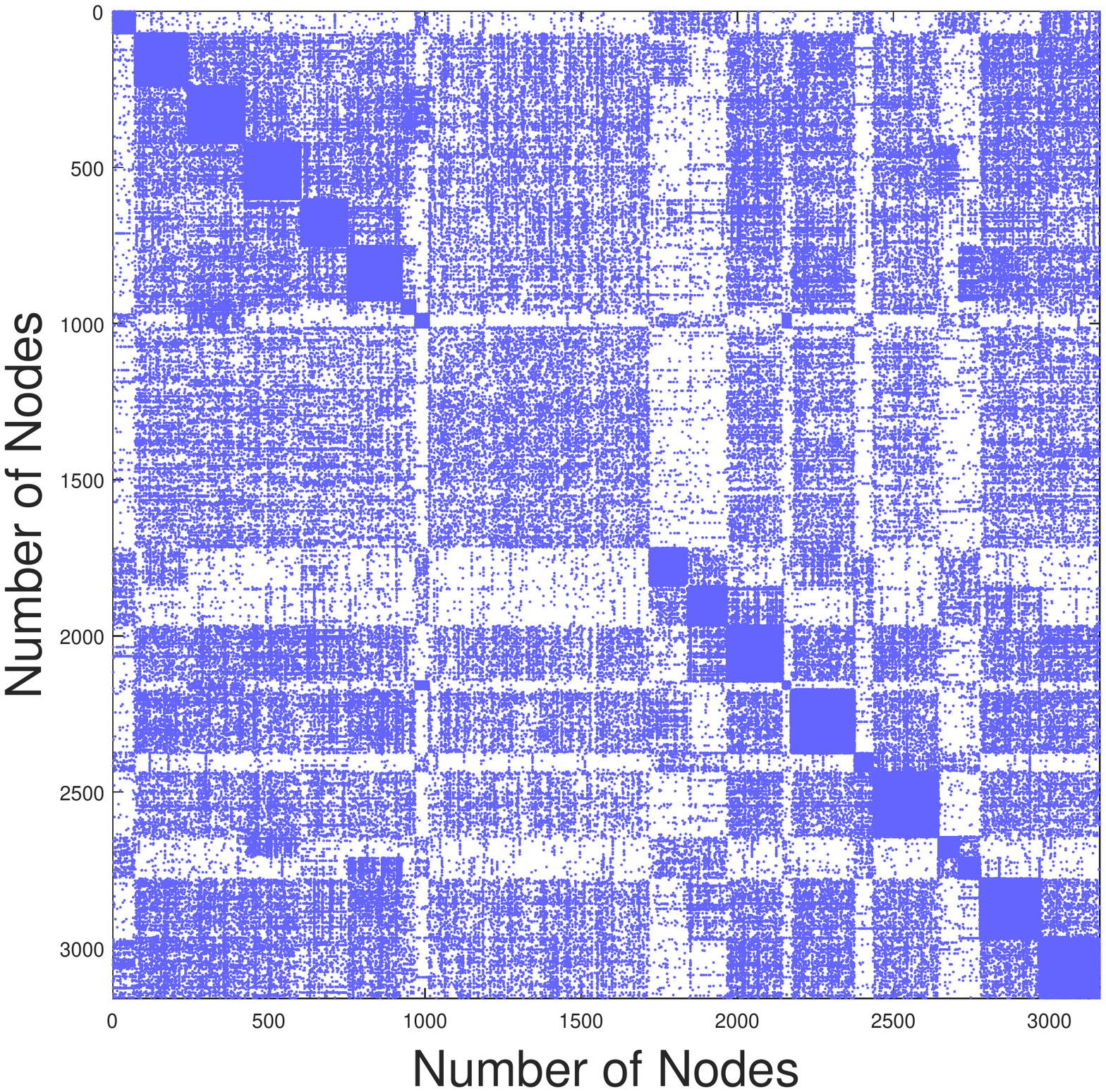,width=0.5\linewidth,clip=} & 
\hspace{40pt}\epsfig{file=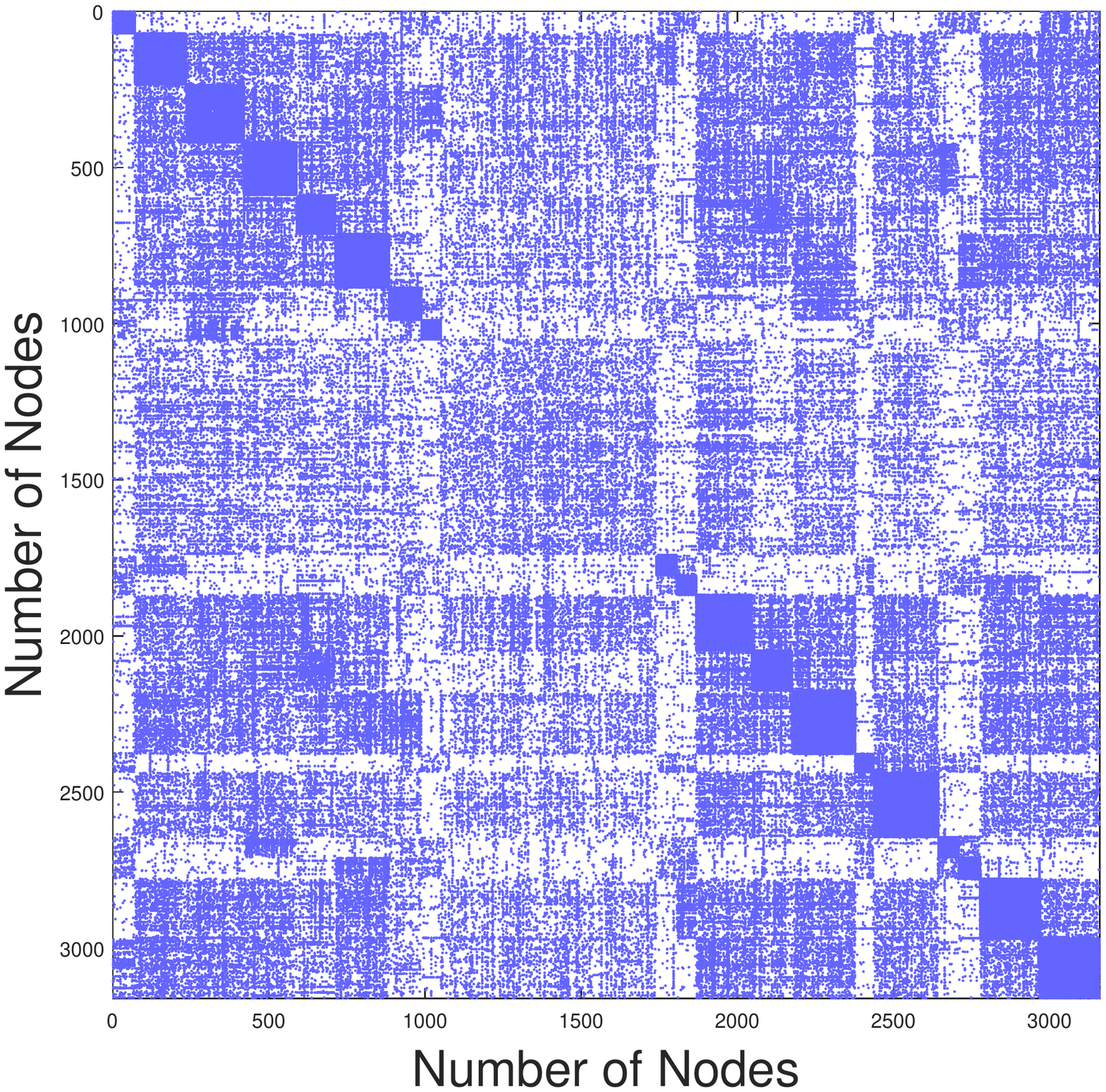,width=0.5\linewidth,clip=} \\
\end{tabular}
\vspace{-5em}
\caption{Community detection plots for parallel MCEM with and without covariate respectively. The two sets of clustering are very similar although the one with covariate (left) appears to be less noisy than the without covariate one (right).}\label{fig:community}
\end{figure}

Recall the K-class SBM with covariates 
\begin{equation}
\label{sbm:covariate:real data}
\log\frac{P_{ij}}{1-P_{ij}}=\theta_{z_iz_j}+\beta^TX(i,j)\mbox{ $i=1,\ldots,n;j=i+1,\ldots,n$ }
\end{equation}
where $P$ is the matrix describing the probability of the edges between any two individuals in the network and the probability of a link between $i$ and $j$ is assumed to be composed of the ``latent'' part given by $\theta_{z_iz_j}$ and the ``covariate'' part given by $\beta^TX(i,j)$ where $\beta$ is a parameter of size $3\times 1$ and $X(i,j)$ a vector of covariates of the same order indicating shared group membership. The vector $\beta$ is implemented here with sum to zero identifiability constraints. 
\\We apply \textbf{Algorithm} \ref{parallel:comm} to fit model~\eqref{sbm:covariate:real data} to the Rice university facebook network with three covariates dorm/house number, gender and class year.
\\We plot the communities found by fitting a SBM without covariates ($\beta=0$ in model (\ref{sbm:covariate})) and a blockmodel with covariates to the given data. Let $\mathcal{C}$ and $\mathcal{C}$ be the two sets of clustering obtained by fitting with and without covariate blockmodel respectively. We define a measure called Minimal Matching Distance (MMD)(\cite{von2010clustering}) to find a \textit{best greedy} 1-1 matching between the two sets of cluster. Suppose $\Pi=\left\{\pi\right\}$ denote the set of all permutations of $k$ labels. Then MMD is defined as
\[\text{MMD}=\frac{1}{n}\min_{\pi\in\Pi}\displaystyle\sum_{i=1}^n 1_{\mathcal{C}(i)\neq\mathcal{C}^{\prime}(i)}\] where $\mathcal{C}(i)$ ($\mathcal{C}^{\prime}(i)$ respectively) denote the clustering label of $i$ in $\mathcal{C}$ ($\mathcal{C}^{\prime}(i)$ respectively). Finding the best permutation then reduces to a problem of maximum bipartite matching and we align the two sets of clustering (with and without covariate) by finding the maximum overlap between the two sets of cluster. The two sets of clustering solutions (with and without covariates respectively) are plotted in Fig. \ref{fig:community}. The estimate of the parameter beta linked with the covariate effects is given by $$\hat{\beta}=[0.7956, -0.1738, -0.6218]^{\prime}$$\\ We compare this finding with the ones observed in \citet{traud2012social}. They studied the ``Facebook" friendships networks of one hundred American institutions at a given point of time. In particular, they calculate the assortativity coefficients and the regression coefficients based on the observed ties to understand homophily at the local level. Further, exploring the community structure reveals the corresponding macroscopic structure. For the Rice University data set, their findings support that residence/dorm number plays a key role in the organization of the friendship network. In fact, residence/dorm number provides the highest assortativity values for the Rice University network. We obtain a similar result, by observing that the effect of the first component of $\hat{\beta}$ is quite high. Further, their study reveals that class year also plays a strong role in influencing the community structure. This is again supported by our finding as the magnitude of the third component in $\hat{\beta}$ is sufficiently large. Finally, as seen in the analysis in \citet{traud2012social}, gender plays a less significant role in the organization of the community structure; a similar conclusion is obtained by examining the magnitude of the second component of $\hat{\beta}$.



Further, we employ the Normalized Mutual Information (NMI) to compare the two sets of clusters. 
The NMI between the two sets of clustering (with and without covariate) is 0.8071 which indicates that the two sets of clustering are quite close i.e. the effects of the covariates in clustering the individuals into groups are not strong.

\section{Conclusion}
Large heterogenous network data are ubiquitous in many application domains. The SBM framework is useful in analyzing networks with a community/group structure. Often, the interest lies in extracting the underlying community structure (inferring about the latent membership vector $z$) in the network, whereas in other situations (where the observed network can be thought of a sample from a large population) the interest lies in the estimation of the model parameters ($(\theta,\beta,\pi)$). There are certainly fast methods (e.g. the pseudo-likelihood based method in \citep{amini2013}) available for community detection in large networks, but these approximations are readily not applicable to settings when there is also covariate information available for the nodes. Further, comparison with some of the existing latent space models with covariates reveal that in certain settings (for sparse networks
and in cases where communities have high interactions) our proposed algorithm performs much better than the existing ones. To obtain maximum likelihood estimates in a large SBM with covariates is computationally challenging. Traditional approaches like MCEM becomes computationally infeasible and hence there is a need for fast computational algorithms. Our proposed algorithm provides a solution in this direction.

The proposed parallel implementation of case-control MCEM  across different cores with communication offers the following advantages: (1) fast computation of the ML estimates of the model parameters by reducing the EM update cost
to $O(Km_0n_0M_r)$ -$Km_0$ being the case-control sample size and $n_0$ the number of subsamples, from $O(n^2M_r)$; (2) the parallel version with communication also exhibits further benefits over its non-communication counterpart, since it provides a bias reduction of the final estimates. It is evident from the results in Section \ref{simresults} that the communications based variant performs much better than the non-communication one when compared to the MCEM on the full data.

\section{Supplementary Materials}
\begin{description}
\item\hspace{20pt} We provide the Matlab codes for the simulations and the real data analysis in the supplementary materials. The Rice University dataset is also provided there. We also provide additional two figures- (a) degree distribution of the Rice University network and (b) a plot of the estimated class probabilities for the covariate model inside the supplementary material. All the matrials are zipped into a file named $\verb supp_materials.zip $. This file include a detailed readme file that describes the contents and instructs the reader on their use. The readme file also contain diagrammatic representations of the two parallel algorithms. All the supplementary files are contained in a single archive and can be obtained via a single download.
\end{description}

\bibliographystyle{Chicago}

\bibliography{Reference_comm}
\end{document}